\documentclass[twocolumn,
	aps, prd,
	10pt, notitlepage, %a4paper,
        floats, floatfix,
	amsmath, amssymb, amsfonts, eqsecnum,
	superscriptaddress,
	showpacs, showkeys,
	nofootinbib,
 	longbibliography,
]{revtex4-2}

\usepackage{mathtools}
\usepackage{wrapfig}
\usepackage{graphicx} % include figures
\usepackage[dvipsnames]{xcolor}
\usepackage{xspace} % Sensible space treatment at end of simple macros
\usepackage{bm} % bold math
\usepackage{amsmath,amsfonts,amssymb,amsthm,mathtools}
\usepackage[utf8]{inputenc} % for some references from inspirehep.net
\usepackage{multirow}
\usepackage{array}
\xdefinecolor{mylinkcolor}{rgb}{0,0,0.5}
\usepackage[
	bookmarksnumbered, bookmarksopen, bookmarksopenlevel=2,
	breaklinks=true,
	colorlinks=true, filecolor=mylinkcolor, citecolor=mylinkcolor,
	linkcolor=mylinkcolor, urlcolor=mylinkcolor, menucolor=mylinkcolor,
]{hyperref}

\usepackage{verbatim}
\usepackage{mathrsfs}
\usepackage{color}
\usepackage{enumitem}
\usepackage{comment}
\usepackage{tikz}
\usepackage{tikz-3dplot}
\usepackage{mathbbol}
\usepackage{bbm}
\usepackage{placeins}

\usetikzlibrary{decorations.pathmorphing}
\usetikzlibrary{arrows}

\tikzset{snake it/.style={decorate, decoration=snake}}

\usepackage[normalem]{ulem}
\newcommand{\be}{\begin{equation}}
\newcommand{\ee}{\end{equation}}
\newcommand{\bse}{\begin{subequations}}
\newcommand{\ese}{\end{subequations}}

\newcommand{\bpm}{\begin{pmatrix}}
\newcommand{\epm}{\end{pmatrix}}

\usepackage{colortbl}
\definecolor{blue2}{cmyk}{1, 0.1, 0.1, 0}

\definecolor{pyBlue}{RGB}{31, 119, 180}
\definecolor{pyRed}{RGB}{214, 39, 40}
\definecolor{pyGreen}{RGB}{44, 160, 44}
\definecolor{pyBlue2}{RGB}{0, 111, 237}
\definecolor{pyRed2}{RGB}{224, 52, 36}

\usepackage{colortbl}
\definecolor{summersky}{cmyk}{0.71,0.33,0,0.5}
\definecolor{flamingo}{cmyk}{0,0.51,0.71,0.5}
\definecolor{rp}{cmyk}{0.2, 1, 0.6, 0}
\definecolor{pacificblue}{cmyk}{0.95,0.3,0, 0.5}
\definecolor{gray60}{cmyk}{0.4,0.4,0,0.8}

\newcommand{\dd}{\mathop{\mathrm{d}\!}{}}

\begin{document}

\title{Modeling Relativistic Tidal Disruptions of MESA Stars}
\author{Liam M. Wang}
\email{lwang959@wisc.edu}
\affiliation{Department of Astrophysical Sciences, Princeton University, Princeton, NJ 08544, USA}
\affiliation{Department of Astronomy, University of Wisconsin--Madison, Madison, WI 53706, USA}%

\author{Giovanni Maria Tomaselli}
\email{tomaselli@ias.edu}
\affiliation{School of Natural Sciences, Institute for Advanced Study, Princeton, NJ 08540, USA}%

\author{Zihan Zhou}
\email{zihanz@princeton.edu}
\affiliation{Department of Physics, Princeton University, Princeton, NJ 08544, USA}%

\begin{abstract}
Tidal disruption events (TDEs) occur when a star passes so close to a black hole that its self-gravity is overcome by the external tidal field. As the star passes, it initially deforms, then is ripped apart, and some of its material eventually falls back on bound orbits, forming an accretion disk around the black hole. A Newtonian model of TDEs, based on stellar perturbation theory of MESA stars, was recently introduced as an alternative to computationally intensive hydrodynamical simulations. In this work, we add relativistic corrections to the model, incorporating equatorial Kerr geodesics, relativistic tidal fields, and relativistic fallback times. Compared to the Newtonian case, we find that stars are disrupted earlier in their orbit, which gives them less time to accumulate physical deformations. Additionally, we find that the increased distance from the black hole at the time of disruption makes the fallback time longer. However, the black hole spin has a negligible impact on fallback time, except for orbits with exceptionally close pericenter. Our results allow for a more accurate calculation of fallback rates than the Newtonian model, while also remaining computationally cheap. The code is available on GitHub.

\end{abstract}

\maketitle

\section{Introduction}
When a star passes close enough to a massive black hole (BH), the BH's tidal field can overcome the star's self-gravity and tear it apart in a tidal disruption event (TDE) \cite{Rees1998}. About half of the resulting debris remains bound to the BH and gradually falls back onto it, powering a luminous flare whose light curve depends on the mass fallback rate $\dd M/\dd t$ \cite{1990ApJ...351...38C, 2011MNRAS.410..359L}. TDEs are now routinely discovered across the electromagnetic spectrum \cite{2021ARA&A..59...21G, 2021ApJ...908....4V}, and their light curves encode properties of the system: the BH mass and spin, the internal structure of the disrupted star, and the pericenter distance of the orbit \cite{2013ApJ...767...25G, Ryu_2020a, Ryu_2020b}.

Inferring these properties from observations requires modeling both the disruption of the star and the ensuing fallback, a problem made difficult by a large separation of scales---the stellar radius is far smaller than the tidal radius, which for a $1\,M_\odot$ star around a $10^6\,M_\odot$ BH is $\sim\!\!\!20$ Schwarzschild radii. Resolving this hierarchy with hydrodynamic simulations is computationally expensive \cite{Gafton2019,2013ApJ...767...25G, Ryu_2020a, Ryu_2020b,Tejeda:2017cuh}. The cost is even greater for \emph{relativistic} disruptions, which occur around more massive, rapidly spinning BHs, or on deeply plunging orbits---for which the star is disrupted close to the horizon, where general-relativistic effects such as stronger tidal fields and frame dragging become significant. For this reason, the relativistic corrections to TDE observables have not yet been mapped out systematically.

Recently, Ref.~\cite{Zhou2025} introduced a perturbative two-stage model that is far cheaper than hydrodynamic simulations, while still providing more accurate results than traditional analytical arguments. In the first stage, the star is described as a set of stellar oscillation modes, computed with \texttt{GYRE} \citep{Townsend2013} for MESA \citep{Paxton2011} middle-age main sequence stellar profiles. These modes are then driven by the time-dependent tidal field of the BH, and the star is deemed disrupted once the tidal energy it absorbs exceeds its gravitational binding energy. In the second stage, the debris is evolved as a collection of free-falling particles to obtain its energy distribution and the fallback rate. Crucially, the orbit and the tidal field enter this framework only as external inputs, which makes it flexible: relativistic physics can be incorporated simply by replacing the Newtonian orbit and tidal tensor with their Kerr counterparts. In this work we carry out precisely this extension, generalizing the model of \cite{Zhou2025}---originally formulated for equatorial, marginally bound Newtonian orbits---to equatorial geodesics in the Kerr geometry.

Applying the relativistic model, we find that relativistic tides are stronger than in the Newtonian case, so a star accumulates tidal energy more rapidly and is disrupted earlier along its trajectory toward the BH. Since it has less time to deform, the star is less distorted at disruption; this narrows the specific energy distribution of the debris and shifts the fallback rate toward longer times, an effect that grows as the pericenter distance decreases. At large pericenter, the results reduce to the Newtonian predictions of \cite{Zhou2025}. The BH spin enters at higher order in $M/R_p$ and has a comparatively small direct effect on the fallback rate: its main role is to set the minimum pericenter for a non-plunging orbit, which governs the behavior of the most relativistic, close-in encounters.

The code for the Newtonian model from \cite{Zhou2025} is available at \cite{PerTDE:github}, while the relativistic extension developed in this project can be found at
\cite{PerTDE_rel:github}.

\section{Relativistic Tidal Disruptions}
\label{sec:relativistic-tde}

In this section, we generalize the two-stage Newtonian model developed in Ref.~\cite{Zhou2025} to a fully relativistic framework. We begin by describing the relativistic orbits in \S\ref{kerr_geo}. Then, in \S\ref{rel_tides}, we lay down the framework of relativistic tidal interactions between the star and the BH. Finally, in \S\ref{rel_fallback} we calculate the physical displacement and the fallback rate of the stellar fluid elements.

\subsection{Kerr Geodesics}
\label{kerr_geo}
The spacetime around the BH is given by the Kerr metric, which in Boyer-Lindquist coordinates (with $G=c=1$) reads
\begin{equation}
\begin{split}
ds^{2}
={}& {-\bigg(1 - \frac{2Mr}{\Sigma}\bigg)} \dd t^{2}
- \frac{4Mar \sin^{2}\theta}{\Sigma} \dd t \dd \phi \\
&+ \frac{\Sigma}{\Delta}\dd r^{2} + \Sigma\dd\theta^{2}\\
&+ \bigg(r^{2}+a^{2} + \frac{2Ma^{2} r \sin^{2}\theta}{\Sigma}\bigg)\sin^{2}\theta\dd\phi^{2}\,,
\end{split}
\label{eq:kerr_metric}
\end{equation}
where $a$ is the BH spin, and $\Sigma$ and $\Delta$ are defined as
\begin{align}
\Sigma &= r^{2} + a^{2}\cos^{2}\theta\,, \\
\Delta &= r^{2} - 2Mr + a^{2}\,.
\end{align}

A test particle in the Kerr spacetime follows a geodesic determined by three conserved quantities: the energy $E$, the axial angular momentum $L_z$ per unit mass, and the Carter constant $\mathcal{Q}$. The geodesic equations are
\begin{align}
\label{eq:dtdtau}
\Sigma \frac{\dd t}{\dd\tau}
&= -a\left(aE\sin^{2}\theta - L_z\right) + \frac{r^{2}+a^{2}}{\Delta}\,P(r)\,,
\\[6pt]
\label{eq:drdtau}
\Sigma \frac{\dd r}{\dd\tau}
&= \pm\sqrt{P(r)^{2} - \Delta\left(r^{2} + (L_z - aE)^{2} + \mathcal{Q}\right)}\,,
\\[6pt]
\label{eq:dthetadtau}
\Sigma \frac{\dd\theta}{\dd\tau}
&= \pm\sqrt{\mathcal{Q} - \cos^{2}\theta\left(a^{2}(1-E^{2}) + \frac{L_z^{2}}{\sin^{2}\theta}\right)}\,,
\\[6pt]
\Sigma \frac{\dd\phi}{\dd\tau}
&= -\left(aE - \frac{L_z}{\sin^{2}\theta}\right) + \frac{a}{\Delta}\,P(r)\,,
\end{align}
where $\tau$ is the proper time and
\begin{equation}
P(r) \equiv E\left(r^{2} + a^{2}\right) - a L_z\,.
\end{equation}
The $\pm$ signs in the radial and polar equations select the two branches (outbound and inbound w.r.t.\ $R_p$) of the $r$- and $\theta$-motion; each flips at its corresponding turning point where the expression under the square root vanishes.

Constructing the local frame in which the tidal field is evaluated (see \S\ref{rel_fallback}) additionally requires the Marck tetrad rotation angle $\psi$ \citep{Marck1983}, which parametrizes the rotation of the parallel-transported orthonormal tetrad along the orbit. It obeys
\begin{equation}
\Sigma \frac{\dd\psi}{\dd\tau}
= \sqrt{\mathcal{K}}\left(
\frac{P(r)}{\mathcal{K} + r^{2}}
+ a\,\frac{L_z - aE\sin^{2}\theta}{\mathcal{K} - a^{2}\cos^{2}\theta}
\right)\,,
\label{eq:geodesics_psi}
\end{equation}
where $\mathcal{K} = \mathcal{Q} + (L_z - aE)^{2}$ is a constant associated with the Killing tensor of the Kerr geometry.

In this work, we focus on stars on equatorial orbits, where $\theta=\pi/2$. In this case the Carter constant vanishes, $\mathcal{Q}=0$, the polar motion is frozen ($\dd\theta/\dd\tau=0$ since $\theta\equiv\pi/2$) and $\mathcal{K}=(L_z - aE)^{2}$, while the remaining equations simplify with $\sin^{2}\theta=1$ and $\cos^{2}\theta=0$.

We assume that the star is on a marginally bound orbit, and correspondingly set $E=1$. The axial angular momentum can be expressed as a function of the pericenter distance $R_p$ as
\begin{equation}
    L_z = \frac{\sqrt{2M R_p\,\Delta(R_p)} - 2Ma}{R_p - 2M}\,,
\label{eq:momentum}
\end{equation}
where $\Delta(R_p) = R_p^{2} - 2M R_p + a^{2}$ is the metric function $\Delta$ evaluated at the pericenter. By convention, we keep $L_z>0$ and change the sign of the BH spin to flip the direction of the orbit, with $a>0$ ($a<0$) corresponding to prograde (retrograde) orbits, and $a=0$ to Schwarzschild geodesics. Furthermore, we set the proper time to $\tau=0$ when $r=R_p$. We show examples of relativistic orbits in the leftmost panels of Fig.~\ref{fig:density_mass}.

\subsection{Relativistic Tides}
\label{rel_tides}
Following \citep{Zhou2025, PhysRevD.73.104029, PhysRevD.73.104030}, we employ an EFT-inspired approach, where the star is modeled as a point particle dressed with multipole moments,
\begin{equation}
    S=\int\dd\tau\bigg[-M_\star-\frac12Q_{ij}E^{ij}+\mathcal L_Q(Q_{ij},\dot Q_{ij})+\ldots\bigg]\,,
\label{rel_action}
\end{equation}
where $M_\star$ is the star mass, $Q_{ij}$ is its quadrupole moment, $E_{ij}$ is the electric part of the tidal field, and higher-multipole terms are neglected. The term $\mathcal L_Q(Q_{ij},\dot Q_{ij})$ encodes the internal degrees of freedom of the star, which we model as a collection of harmonic oscillators. We compute the eigenmodes and eigenfrequencies for MESA stars using \texttt{GYRE} \citep{Townsend2013}.

In the equatorial plane, the first and third spatial legs of the Marck tetrad span the orbital plane, while the second leg is normal to it. We therefore use the local Cartesian labels $(x,y,z)=(1,3,2)$, so that $z$ is normal to the orbit. Throughout this first stage of the calculation, these labels refer to the parallel-transported Marck frame. In this frame, the nonzero components of the tidal field of the Kerr BH, expressed in terms of the rotation angle $\psi$ of \eqref{eq:geodesics_psi}, are~\cite{Kesden2012a}
\begin{align}
E_{xx} &= \frac{M}{r^{3}}\left(1 - 3\cos^{2}\psi\,\frac{r^{2} + (L_z - a)^{2}}{r^{2}}\right)\,,
\label{eq:tidal_xx}
\\[6pt]
E_{xy} &= E_{yx} = -\frac{3M\sin\psi\cos\psi\,\bigl[r^{2} + (L_z - a)^{2}\bigr]}{r^{5}}\,,
\label{eq:tidal_xy}
\\[6pt]
E_{yy} &= \frac{M}{r^{3}}\left(1 - 3\sin^{2}\psi\,\frac{r^{2} + (L_z - a)^{2}}{r^{2}}\right)\,,
\label{eq:tidal_yy}
\\[6pt]
E_{zz} &= \frac{M}{r^{3}}\left(1 + \frac{3(L_z - a)^{2}}{r^{2}}\right)\,.
\label{eq:tidal_zz}
\end{align}
Additionally, the tidal field acts as a source in the equations of motion for the quadrupole moment,
\begin{equation}
    \frac{d}{d\tau} \frac{\partial}{\partial \dot Q_{ij}} \mathcal{L}_Q - \frac{\partial \mathcal{L}_Q} {\partial Q_{ij}} = -\frac{1}{2} E_{ij}.
\label{eq:euler-lagrange}
\end{equation}
This can be formally solved using a retarded Green's function $G_\mathrm{ret}$,
\begin{equation}
    Q_{ij}(\tau) = -\frac{1}{2}\int_{-\infty}^{\tau} G_\mathrm{ret}(\tau-\tau ')E_{ij}(\tau ')\dd\tau',
\label{eq:greens_quad}
\end{equation}
as explicitly derived in \citep{Zhou2025}. Furthermore, within linear fluid perturbation theory \citep{PhysRevLett.12.114, Schenk_2001, 10.1111/j.1365-2966.2004.08459.x, PhysRevD.88.084038}, the Lagrangian $\mathcal L$ can be written as
\begin{equation}
    \mathcal{L} = \sum_{n \ell m}\bigg[\frac{M_\star R_\star^2 \mathcal{N}_{n \ell}}{2} \left(\dot{q}_{n \ell m}^2 - \omega_{n \ell}^2 q_{n \ell m}^2\right) + q_{n \ell m} F_{ n \ell m} \bigg],
\label{eq:lagrangian}
\end{equation}
where $R_\star$ is the star's radius, $\omega_{n \ell}$ are the eigenfrequencies of the stellar oscillations, $\mathcal{N}_{n \ell}$ is a normalization constant, and the driving force $F_{ n \ell m}$ encodes the gravitational coupling between the density perturbations and the background potential.

The total amount of tidal energy inside the star at a given point of its trajectory is
\begin{equation}
    E_Q=-\frac{1}{2}\int_{-\infty}^{\tau}\dot{Q}_{ij}(\tau')\,E^{ij}(\tau')\dd\tau'\,.
\label{eq:tidal_energy}
\end{equation}
This quantity is key to the definition of our disruption criterion. When the tidal energy becomes comparable to the binding energy of the star
\begin{equation}
    E_Q = \gamma|U_\mathrm{bind}|\,,
\label{eq:tde_crit}
\end{equation}
we assume that the star is disrupted, and transition into the second stage of the model, where the fluid elements are treated as noninteracting. The coefficient $\gamma$ is calibrated by comparing the critical disruption distance to hydrodynamical simulations, as done in~\cite{Zhou2025}.

\subsection{Relativistic Fallback Rate}
\label{rel_fallback}

After disruption, we model each fluid element as a freely falling particle. Given the orbital energy $E$ of a fluid element, one can obtain the mass fallback rate $\dd M/\dd T$ by sampling the energy distribution $\dd M/\dd E$ of the perturbed star. In Newtonian gravity, this transformation is given by the Keplerian relation
\begin{equation}
    E=-\frac{1}{2}\bigg(\frac{2\pi GM}{T}\bigg)^{2/3}\,,
\label{eq:E}
\end{equation}
while $E$ is immediately found from the absolute position of the fluid element within the BH potential.

On relativistic orbits, the procedure is more involved. We use Fermi normal coordinates relative to the center of the star and distinguish two closely related orthonormal tetrads. 

Let $\lambda_A^\mu$ denote the parallel-transported Marck tetrad used above, where $A\in\{0,x,y,z\}$ is a tetrad index, $\mu\in\{t,r,\theta,\phi\}$ is a Boyer--Lindquist coordinate index, and $\lambda_0^\mu$ is the four-velocity of the stellar center. Let $\widetilde{\lambda}_A^\mu$ denote the auxiliary tetrad whose Boyer--Lindquist components are given explicitly in Appendix B of Ref.~\cite{Kesden2012b}. Its spatial legs are ordered as $(x,y,z)=(1,3,2)$. The two in-plane legs of this auxiliary tetrad must be rotated by the Marck angle $\psi$ to obtain the parallel-transported tetrad. To make this relation explicit, define
\begin{equation*}
\mathsf{R}(\psi)=
\begin{pmatrix}
\cos\psi & \sin\psi & 0\\
-\sin\psi & \cos\psi & 0\\
0 & 0 & 1
\end{pmatrix}\,.
\end{equation*}
The spatial tetrads are related by
\begin{equation*}
\lambda^\mu{}_i
=\widetilde{\lambda}^\mu{}_j\,
\mathsf{R}^j{}_i(\psi)\,,
\qquad i,j\in\{x,y,z\}\,.
\end{equation*}
The tidal tensor in Eqs.~\eqref{eq:tidal_xx}--\eqref{eq:tidal_zz}, and therefore the deformed stellar profile, are expressed in the parallel-transported frame. We denote the coordinates of a fluid-element displacement in that frame by $X_{\rm PT}^i$. To contract the same physical displacement with the explicit auxiliary tetrad, we transform its components according to
\begin{equation*}
X_{\rm aux}^j
=\mathsf{R}^j{}_i(\psi)X_{\rm PT}^i\,.
\end{equation*}
The two descriptions give the same Boyer--Lindquist displacement because
\begin{align*}
\delta x^\mu
&=\lambda^\mu{}_i X_{\rm PT}^i\\
&=\widetilde{\lambda}^\mu{}_j
  \mathsf{R}^j{}_i(\psi)X_{\rm PT}^i\\
&=\widetilde{\lambda}^\mu{}_j X_{\rm aux}^j.
\end{align*}
Thus, one may either rotate the tetrad and keep the displacement components fixed, or rotate the displacement components and keep the auxiliary tetrad fixed. Our implementation uses the latter form. This is a change of basis, rather than an additional physical rotation of the star.

Using the parallel-transported form, the energy difference of a fluid element with respect to the geodesic of the original point-like star is
\begin{equation}
    \delta E=-g_{\beta\gamma} \lambda_0^\beta X^i \lambda_i^\alpha \Gamma^\gamma_{\alpha t}\,,
\label{eq:deltaE}
\end{equation}
The displacement components in Eq.~\eqref{eq:deltaE} satisfy $X^i\equiv X_{\rm PT}^i$. Here, $g_{\beta\gamma}$ is the Kerr metric, and $\Gamma^\gamma_{\alpha t}$ are the Christoffel symbols. The explicit expressions of the metric, Christoffel symbols, and auxiliary tetrad can be found in Appendices A and B of Ref.~\cite{Kesden2012b}.

However, knowledge of the energy difference $\delta E$ is not sufficient to determine the fallback time $T$, which we assume equal to the radial period $T_r$ of the particle's geodesic. Following again \cite{Kesden2012b}, we can further compute the correction to the angular momentum as
\begin{equation}
 \delta L_z = g_{\beta \gamma} \lambda_0^\beta X^i \lambda_i^\alpha \Gamma^\gamma_{\alpha \phi}\,.
\label{eq:deltaLz}
\end{equation}
Since we model the orbits of the debris as equatorial, their Carter constant is zero and $\delta\mathcal Q=0$.
% and to the Carter constant,
% \begin{equation}
%  \delta\mathcal Q = \delta\mathcal K - 2(L_z - aE)(\delta L_z - a \delta E)\,,
% \label{eq:deltaQ}
% \end{equation}
% where
% \begin{equation}
%  \delta\mathcal K = 2 X^i [\lambda_0^\alpha \lambda_0^\beta \lambda_i^\gamma \nabla_\gamma(\Sigma l_\alpha n_\beta) - r \lambda_i^r]\,.
% \label{eq:deltaK}
% \end{equation}
%Here, $l_\alpha$ and $n_\beta$ are the principal null vectors found in the Petrov classification of the Weyl tensor \cite{wald1984general,Petrov:2000bs, 1969eisp.book.....P}, explicitly defined in~\cite{Kesden2012b}.

The geodesic of the fluid element can then be computed by integrating \eqref{eq:dtdtau}--\eqref{eq:dthetadtau} replacing $E\to E+\delta E$ and $L_z\to L_z+\delta L_z$.
% , and $\mathcal Q\to \mathcal Q+\delta\mathcal Q$. 
Given that the stellar radius is typically much smaller than the tidal disruption radius \cite{Rees1998,Goicovic_2019}, we neglect the evolution of the polar angle here, setting $\dd\theta/\dd\tau=0$ for all fluid elements. Under this assumption, the fallback time is
\begin{equation}
T_r=2\int_{r_\mathrm{min}}^{r_\mathrm{max}}\frac{\dd t}{\dd r}\dd r=2\int_{r_\mathrm{min}}^{r_\mathrm{max}}\frac{\Sigma\dd t/\dd\tau}{\sqrt{R(r)}}\dd r\,,
\label{eq:Tr_def}
\end{equation}
where $r_\mathrm{min}$ and $r_\mathrm{max}$ are the pericenter and apocenter and $R(r)\equiv(\Sigma\,\dd r/\dd\tau)^2$ can be read off \eqref{eq:drdtau}. Substituting \eqref{eq:dtdtau} with $\theta=\pi/2$ in \eqref{eq:Tr_def}, we get
\begin{equation}
    T_r= 2\int_{r_\mathrm{min}}^{r_\mathrm{max}} \frac{(r^2+a^2)\,P(r)}{\Delta\,\sqrt{R(r)}}\,\dd r + \left(a L_z - a^2E\right)\Lambda_r\,,
    \label{eqn:T_r}
\end{equation}
where
\begin{equation}
\begin{split}
\Lambda_r&=2\int_{r_\mathrm{min}}^{r_\mathrm{max}}\frac{\dd r}{\sqrt{R(r)}}\\
&=\frac{4\,K(k_r)}{\sqrt{(1-E^2)(r_1-r_3)(r_2-r_4)}}\,,
\end{split}
\end{equation}
$K$ is the complete elliptic integral of the first kind \cite{Fujita_2009}, $r_1 > r_2 > r_3 > r_4$ are the four real roots of $R(r)$ (so that $r_1=r_\mathrm{max}$ and $r_2=r_\mathrm{min}$), and
\begin{equation}
k_r = \sqrt{\frac{r_1 - r_2}{r_1 - r_3} \frac{r_3 - r_4}{r_2 - r_4}}\,.
\label{eq:k_r}
\end{equation}
The integral term in \eqref{eqn:T_r} can likewise be reduced to elliptic integrals of the first, second, and third kinds \cite{Fujita_2009}; in practice we evaluate it by numerical quadrature. Thus, we use \eqref{eqn:T_r} to compute the radial period of all fluid elements of the disrupted star. Only bound, non-plunging elements contribute: those with $E+\delta E\geq1$ escape, while those whose perturbed radial potential has no turning point outside the horizon are directly captured by the BH. We normalize $\dd M/\dd T$ to the total stellar mass, so that its integral gives the returning mass fraction rather than unity. The resulting mass distribution $\dd M/\dd T$ for $T_r$ is the desired fallback rate.

\section{Results}

We now describe the results obtained using the TDE model from \S\ref{sec:relativistic-tde}, and compare them with Newtonian ones. The tidal energy results are discussed in \S\ref{sec:tidal_energy}, the stellar density profiles in \S\ref{sec:density}, and the energy distribution and fallback rate in \S\ref{sec:dMdE-dMdT}.

\subsection{Tidal Energy}
\label{sec:tidal_energy}

\begin{figure}
    \centering
    \includegraphics[width=0.48\textwidth]{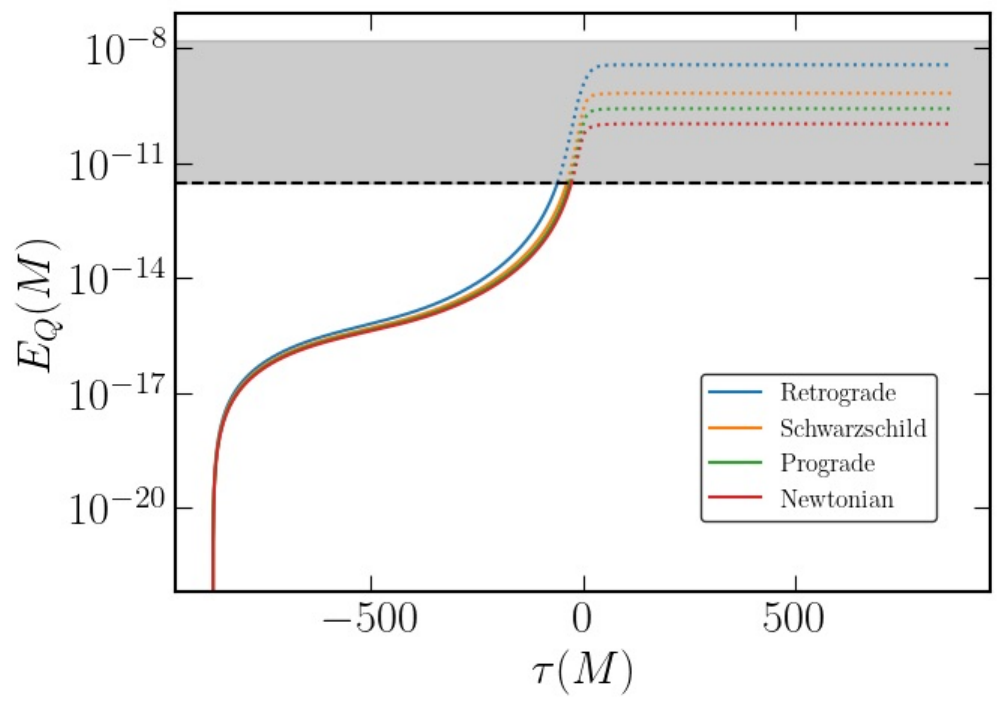}
    \caption{Tidal energy $E_Q$, as defined in \eqref{eq:tidal_energy}, for $M_\star=1\,M_\odot$ as a function of the star's proper time for Newtonian, Schwarzschild, Kerr prograde and Kerr retrograde orbits (with spin $|a|=0.99$). All cases assume the same pericenter distance $R_p=6.2M$. The dashed line marks the threshold \eqref{eq:tde_crit} which defines the point of TDE in our model, and we see that the relativistic cases reach this threshold earlier than the Newtonian case. For purely illustrative purposes, we show how the tidal energy would continue past that point if we didn't transition to the second stage of the model, where fluid elements are treated as free particles.}
    \label{fig:tidal}
\end{figure}

We show in Fig.~\ref{fig:tidal} the tidal energy $E_Q$, as defined in \eqref{eq:tidal_energy}, as a function of the proper time. To emphasize the impact of relativistic corrections, we fix a small pericenter distance $R_p=6.2M$ for $M_\star=1\,M_\odot$, and compare the Newtonian, Schwarzschild, Kerr prograde and Kerr retrograde cases (with maximal spin $a=0.99$).

Relativistic effects increase the amount of tidal energy pumped into the star. Additionally, the BH spin increases (decreases) the tidal energy for retrograde (prograde) orbits. A larger amount of tidal energy means that (1) the value of $R_p$ corresponding to the nondisruption threshold is larger, and (2) at fixed $R_p$, the star is disrupted earlier along its orbit, because the critical condition \eqref{eq:tde_crit} is reached sooner. This is visible in Fig.~\ref{fig:tidal}, where the threshold $\gamma|U_{\text{bind}}|$ is drawn as a black horizontal dashed line.\footnote{Above such critical value of the tidal energy, our model prescribes to transition into the second stage, where fluid elements are treated as free particles. However, for purely illustrative purposes, we continue the calculation of $E_Q$ past that point.}

For the parameters of Fig.~\ref{fig:tidal}, the relativistic tides disrupt the star farther from the BH than in the Newtonian case. Defining $\Delta r$ as the difference between the relativistic and Newtonian disruption radii, we find $\Delta r\approx0.5\text{--}1.8M$ across the three relativistic cases, i.e. $\Delta r/R_p\approx7.4\text{--}29.2\%$ of the pericenter distance. As expected, this correction shrinks as the pericenter distance increases, dropping to $\Delta r/R_p\approx1.1\text{--}4.0\%$ at $R_p=16M$.

\vspace{-15pt}
\subsection{Equatorial Density Profiles}
\vspace{-5pt}
\label{sec:density}

\begin{figure*}[t]
    \centering
    \vspace{-10pt}
    \includegraphics[width=\linewidth]{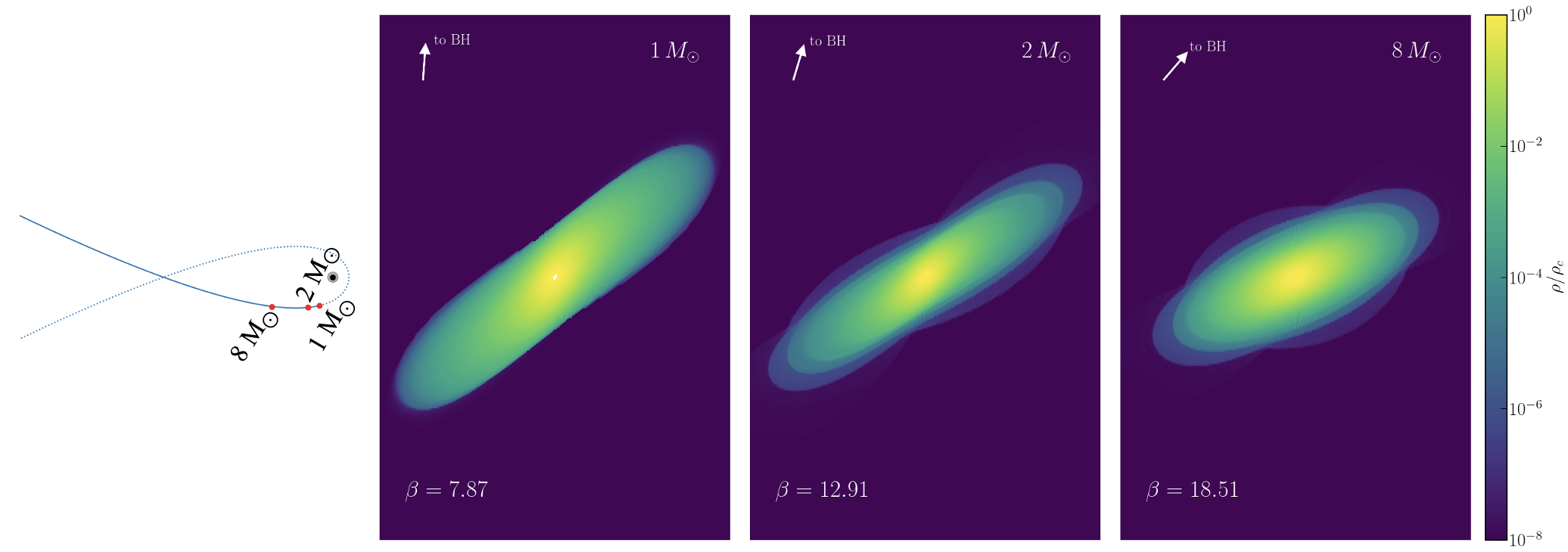}\\[-14pt]
    \includegraphics[width=\linewidth]{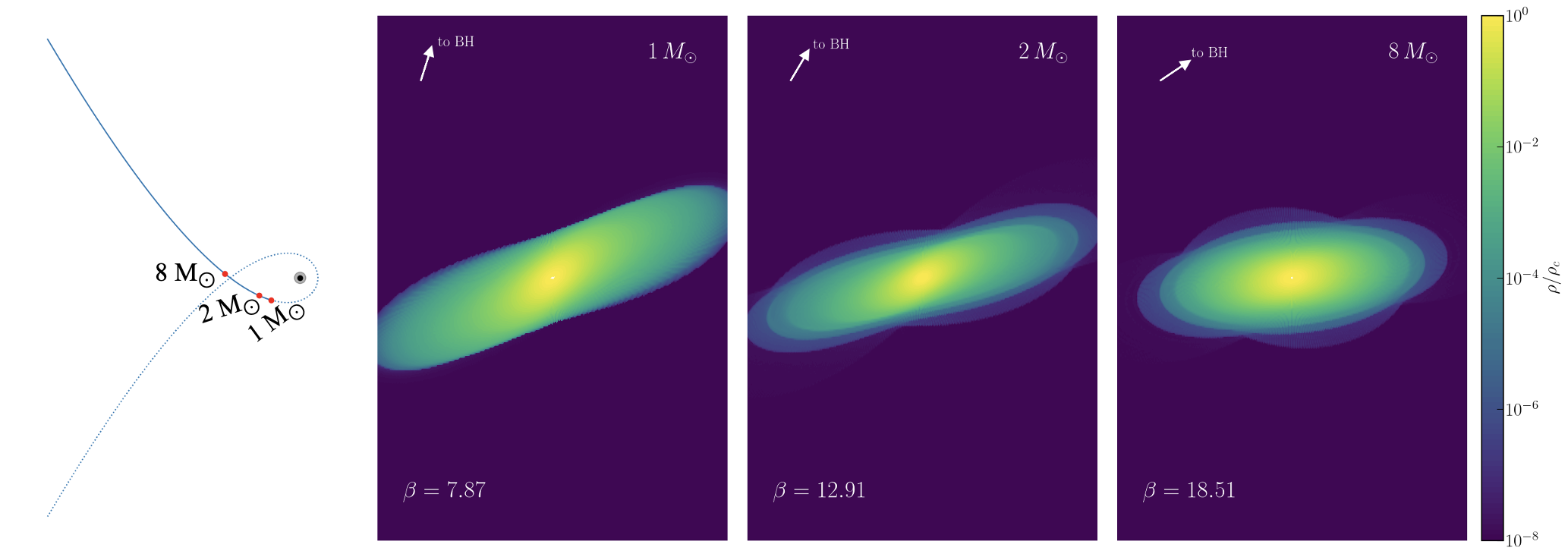}\\[-14pt]
    \includegraphics[width=\linewidth]{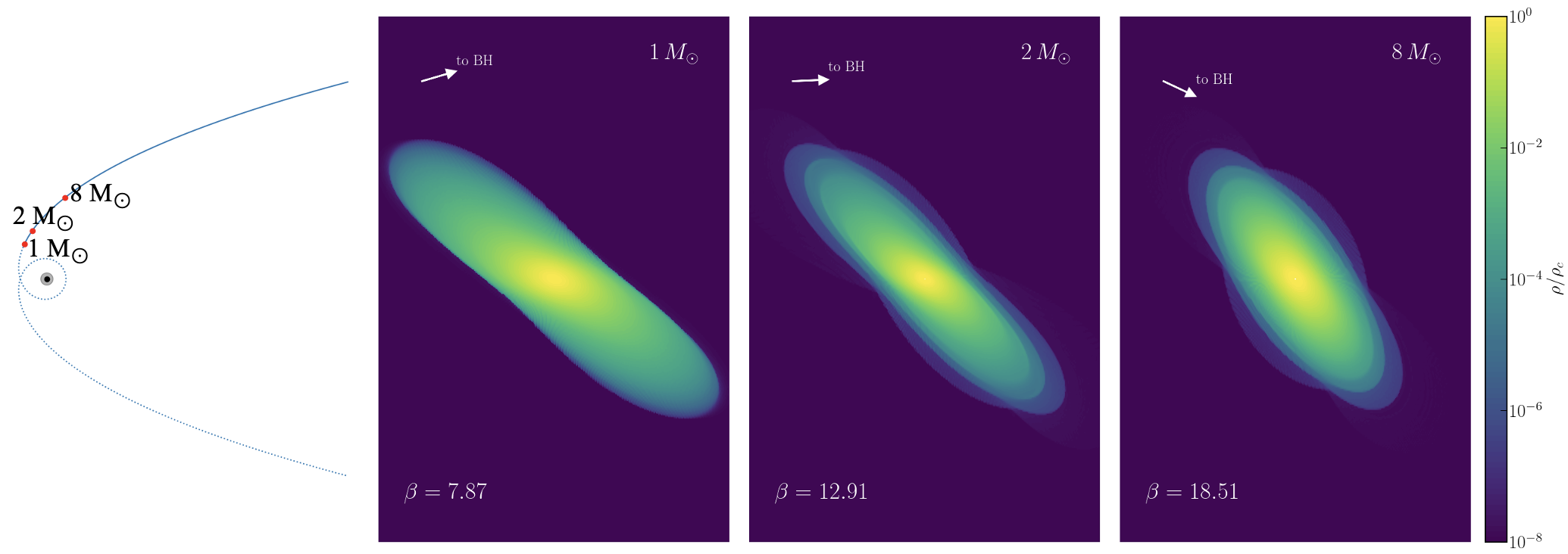}\\[-14pt]
    \includegraphics[width=\linewidth]{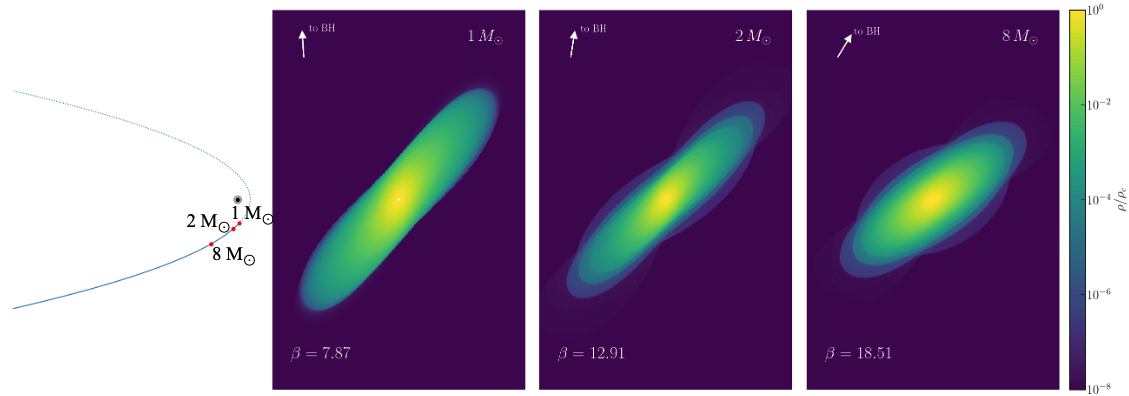}\\[-10pt]
    \caption{\emph{Leftmost panels}: Kerr prograde ($a = 0.99$, \emph{top}), Schwarzschild ($a = 0$, \emph{2nd row}), Kerr retrograde ($a =-0.99$, \emph{3rd row}), and Newtonian (\emph{bottom}) orbits with pericenter $R_p=6.2M$ around a BH with $M=10^6\,M_\odot$. \emph{Middle-left, middle-right and right panels:} deformed equatorial projected column density profiles of stars with $M_\star=1\,M_\odot,2\,M_\odot,8\,M_\odot$ at their TDE time. Their position along the orbit at the TDE time is indicated with red dots in the leftmost panels. The panels also report the values of $\beta=R_\star(M/M_\star)^{1/3}/R_p$.}
    \label{fig:density_mass}
\end{figure*}
\begin{figure*}
    \centering
    \vspace{-10pt}
    \includegraphics[width=\linewidth]{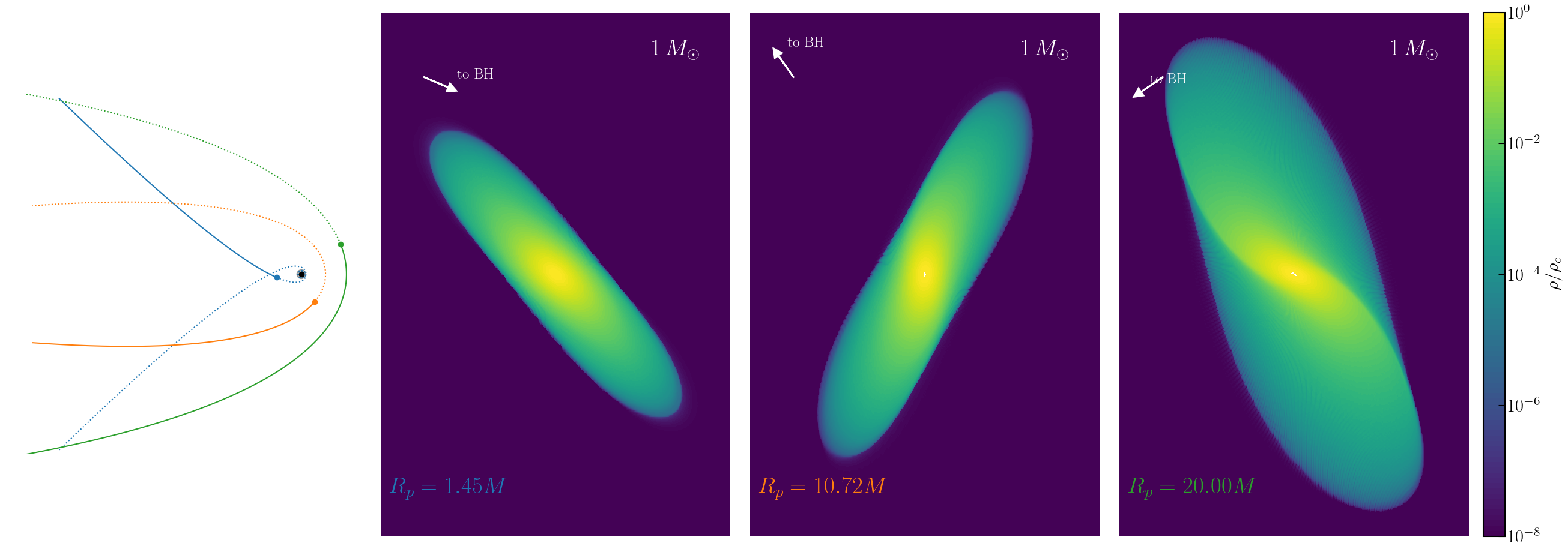}\\[-14pt]
    \includegraphics[width=\linewidth]{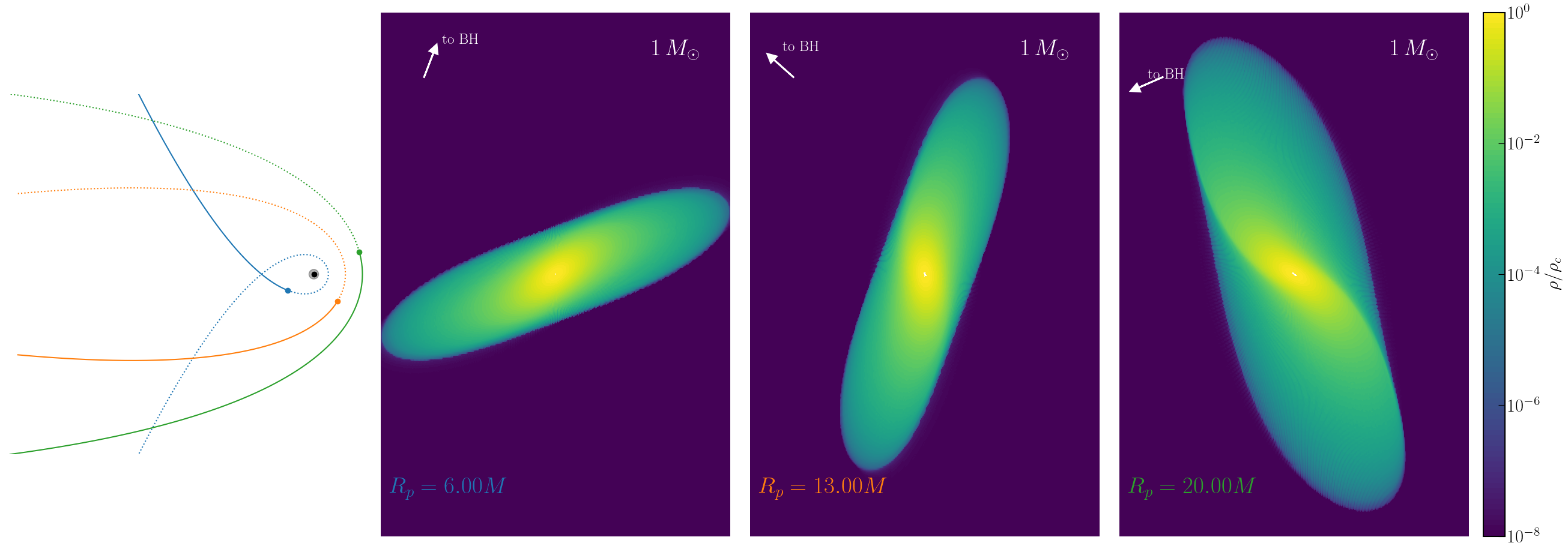}\\[-14pt]
    \includegraphics[width=\linewidth]{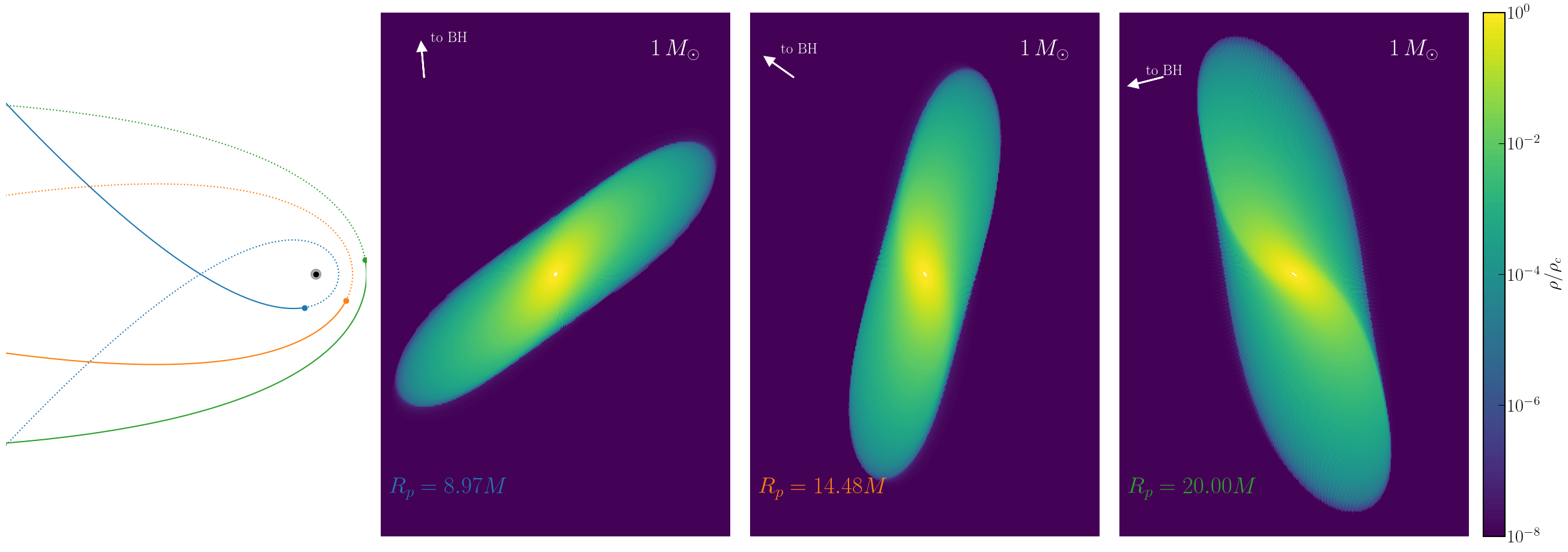}\\[-14pt]
    \includegraphics[width=\linewidth]{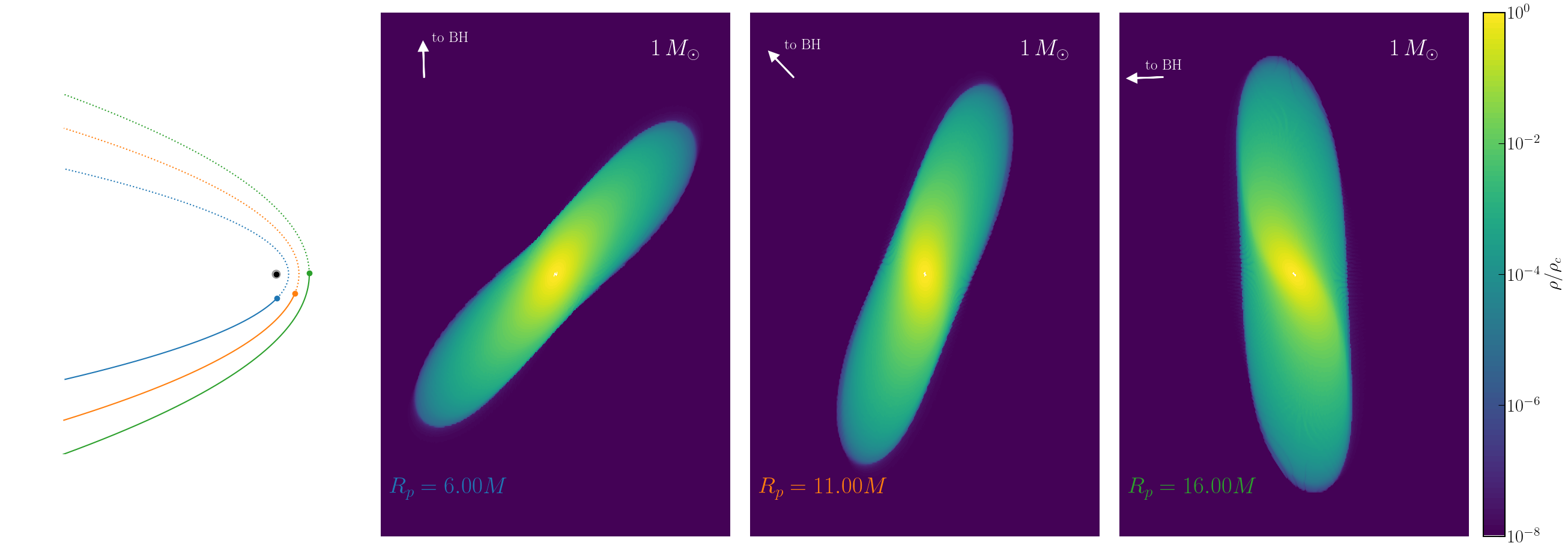}\\[-10pt]
    \caption{Same as Fig.~\ref{fig:density_mass}, but varying the pericenter distance $R_p$ rather than the star mass, which is fixed at $M_\star=1\,M_\odot$. The values of $R_p$ roughly span from the ISCO to $\sim\!84\%$ of nondisruption limit for each trajectory (relativistic: $\sim\!20M$, Newtonian: $\sim\!16M$). The ISCO is used here only as a conventional marker: marginally bound orbits can remain non-plunging even inside it. The dots on the trajectories indicate the positions at TDE time for the various cases.}
    \label{fig:density_rp_full}
\end{figure*}

\begin{figure*}[t]
    \centering
    \includegraphics[width=\linewidth]{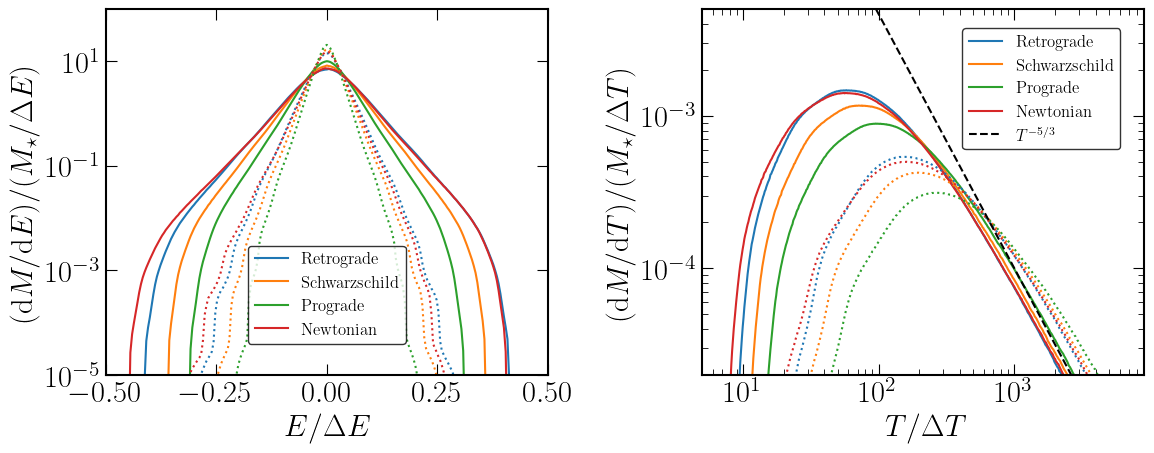}
    \includegraphics[width=\linewidth]{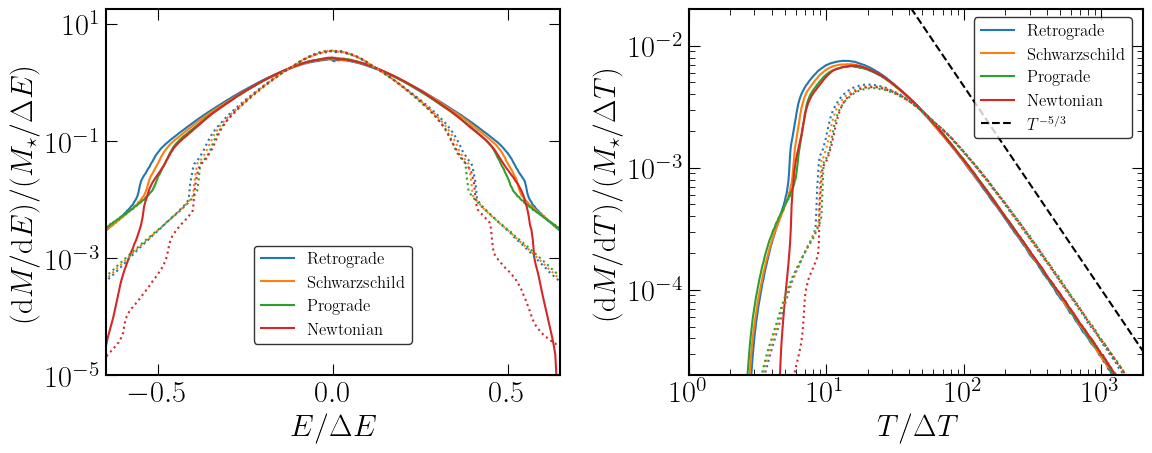}
    \caption{Specific energy distribution and fallback rate for $M_\star=1\,M_\odot$, $M=10^6 M_\odot$, and $R_p$ near the ISCO radius (\emph{top}), which again we pick as a conventional marker, specifically $R_p=6M$ for the Schwarzschild and Newtonian cases; $R_p=5.3M$, $a=0.2$ for Kerr prograde; $R_p=6.6M$, $a=-0.2$ for Kerr retrograde, and $R_p$ around $\sim\!84\%$ of the nondisruption threshold (\emph{bottom}), namely $R_p=16M$ for the Newtonian case and $R_p=20M$ for the relativistic cases with the same corresponding BH spin as the top panels. The dotted lines showcase the same distributions for a $2M_\odot$ star around the same BH; its Newtonian and relativistic cases are both similarly $\sim\!84\%$ ($21M$ and $25M$ respectively) of their nondisruption thresholds.}
    \label{fig:fallback_cases}
\end{figure*}
\begin{figure*}
    \centering
    \includegraphics[width=\linewidth]{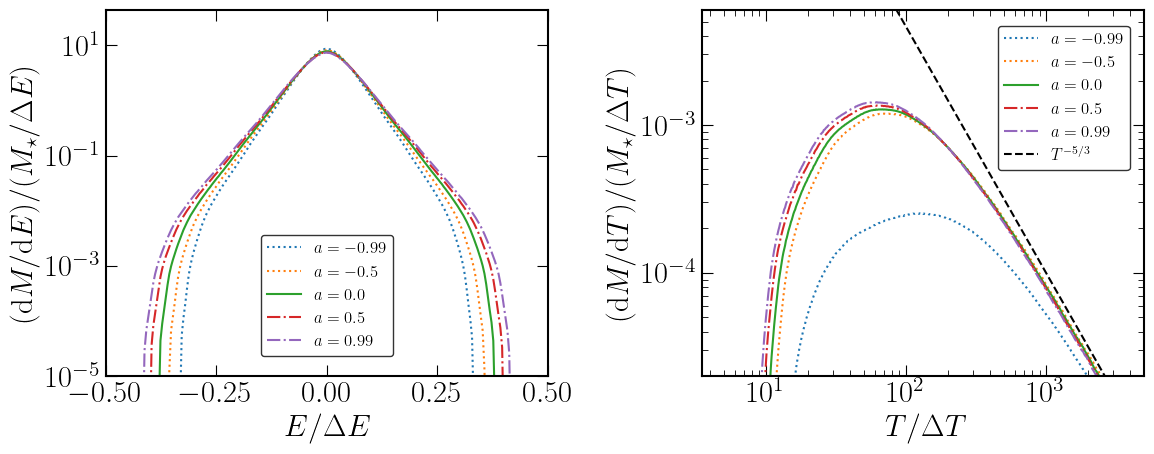}
    \includegraphics[width=\linewidth]{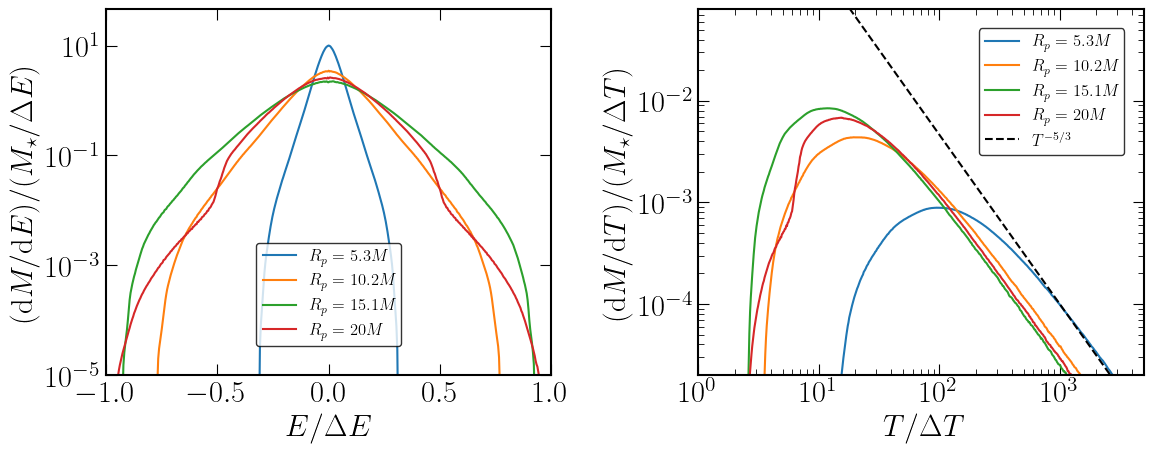}
    \caption{Same as the $1M_\odot$ cases in Fig.~\ref{fig:fallback_cases}, but varying the BH spin at fixed $R_p=6.2M$ (\emph{top}) and varying the pericenter distance at fixed $a=0.2$ (\emph{bottom}). In the top panel, dotted lines denote retrograde orbits while dash-dotted lines denote prograde orbits.}
    \label{fig:fallback_spin_rp}
\end{figure*}

Next, we analyze the equatorial projected column density profiles of the deformed stellar profiles at the moment the TDE condition \eqref{eq:tde_crit} is reached. We show these profiles for Schwarzschild, Kerr prograde and  retrograde orbits in Fig.~\ref{fig:density_mass} (where we vary $M_\star$) and Fig.~\ref{fig:density_rp_full} (where we vary $R_p$). The results can be compared to the Newtonian case, shown in the bottom rows of Fig.~\ref{fig:density_mass} and Fig.~\ref{fig:density_rp_full}.

We first note that smaller $R_p$ and higher $M_\star$ correspond to a lower degree of deformation of the star. While this might initially seem counterintuitive, it is actually due to the fact that in cases where the tidal energy $E_Q$ is higher, the star is disrupted earlier in the orbit, when it has not had the time to deform significantly yet. For the same reason, the effect of relativistic corrections slightly reduces the amount of stellar deformation at TDE, which can be seen by comparing the relativistic and Newtonian cases in Fig.~\ref{fig:density_mass} and Fig.~\ref{fig:density_rp_full}. This effect is bigger for smaller $R_p$, which aligns with the expectation that relativistic corrections enter at order $\mathcal O(M/R_p)$. The profiles are also rotated differently in the two cases because the disruption occurs at different points along the orbit, on which the direction pointing from the star to the BH depends.

The effect of the BH spin is also enhanced when the pericenter distance is small, but this is a higher order effect in $M/R_p$ and it is thus hard to see by eye. At fixed $R_p$, retrograde orbits lead to a slightly reduced deformation at TDE, in addition to the rotation effect described earlier.

\vspace{-15pt}
\subsection{Energy Distribution and Fallback Rate}
\vspace{-5pt}
\label{sec:dMdE-dMdT}

Figures~\ref{fig:fallback_cases} and~\ref{fig:fallback_spin_rp} show the energy distribution $\dd M/\dd E$ and fallback rate $\dd M/\dd T$, varying the pericenter distance and the BH spin. Following conventions in the TDE literature, we normalize specific energies and fallback times by
\begin{equation}
    \Delta E=\frac{MR_\star}{R_p^2}\,,\qquad\Delta T=\frac{M}{\Delta E^{3/2}}\,.
\end{equation}

As expected, at large $R_p$ the relativistic results recover the Newtonian limit from \cite{Zhou2025}. To properly showcase this in the bottom panels of Fig.~\ref{fig:fallback_cases}, for each $M_\star$ we choose $R_p$ to be a similar percentage ($\sim\!84\%$) of the respective nondisruption threshold for the Newtonian and relativistic cases. This means that for $M_\star=1M_\odot$ we pick $R_p=16M$ for the Newtonian case and $R_p=20M$ for the relativistic cases, while for $M_\star=2M_\odot$ we pick $R_p=21M$ for the Newtonian case and $R_p=25M$ for the relativistic cases.

For pericenter distances as small as the ISCO of each case (top panels of Fig.~\ref{fig:fallback_cases}), the relativistic and spin effects become much more pronounced. The energy distributions are narrower here because the star is less deformed at disruption, and the fallback times increase accordingly. The wings of the energy distribution, and thus the rising portion of the fallback curve, show some kinks and discontinuities for more massive stars, e.g., in the bottom right panel of Fig.~\ref{fig:fallback_cases}. These are due to jumps in the density profiles of the envelope, and have also been observed in \cite{Zhou2025, Golightly_2019}.

From the top panels of Fig.~\ref{fig:fallback_spin_rp}, we see that, even on very relativistic orbits, changing the spin does not affect the fallback rate as much as changing $R_p$. Only the near-extremal case $a=-0.99$ features a significant shift of the fallback rate curve to longer times, but that is because the fixed $R_p=6.2M$ is approaching the plunge boundary at that spin. Its normalization is also reduced by a factor $\simeq2.2$: at this spin $27\%$ of the star's mass---more than half of the bound debris---plunges into the BH instead of returning.

These findings support the conclusion that spin does not contribute as much to relativistic corrections as $R_p$ does, because it enters at a higher order in $M/R_p$. However, spin does affect how close a star can get to a supermassive BH without plunging, which in turn influences the behavior of very close orbits.

\vspace{-12pt}
\section{Conclusions}
\vspace{-5pt}
In much of the previous literature, modeling of TDEs has relied either on computationally expensive hydrodynamical simulations, or crude semi-analytical estimates. Reference~\cite{Zhou2025} introduced a two-stage Newtonian model that can offer more accurate predictions than analytical arguments, while also remaining computationally inexpensive.

In this work, we extend that model to account for relativistic corrections, with the goal of further increasing its accuracy. To do this, we considered stars on equatorial orbits in the Kerr geometry, and found that relativistic effects make tidal forces stronger, leading to a quicker accumulation of tidal energy in the star and to an earlier breakup. We found spin effects to be generally small, only producing significant corrections to the fallback rate for stars on orbits with very close pericenter.

Other extensions are possible. These include further generalizations such as non-equatorial orbits, improvements of the TDE model, such as incorporating stellar mode coupling in the first stage, or debris self-gravity in the second stage. We hope to return to these ideas in future works.

\section*{Acknowledgments}

We thank Romain Teyssier and Eliot Quataert for helpful discussions. G.M.T. gratefully acknowledges support from the Rubicon Fellowship, awarded by the Netherlands Organisation for Scientific Research (NWO), Grant ID \href{https://doi.org/10.61686/WYKDB06497}{https://doi.org/10.61686/WYKDB06497}. 

\bibliography{bibliography}
\end{document}